\documentclass[]{raa}
\usepackage{graphicx,times}
\usepackage{natbib}
\usepackage{amssymb,amsmath}
\bibpunct{(}{)}{;}{a}{}{,}

\usepackage[a4paper=true,dvipdfm=true,pagebackref=true]{hyperref}
\hypersetup{pdftitle = The title of my PDF, pdfauthor = My name, pdfsubject= The subject, pdfkeywords = keyword1 keyword2 keyword3}
\hypersetup{colorlinks = true, linkcolor = green, anchorcolor = red, citecolor = blue, filecolor = red, pagecolor = red, urlcolor = red}

\usepackage{natbib}
\usepackage{graphicx}
\usepackage{multirow}
\usepackage{txfonts}

\begin{document}

\title{Accretion Flow Properties of XTE J1118+480 During Its 2005 Outburst
}

% \volnopage{ {\bf 20XX} Vol.\ {\bf X} No. {\bf XX}, 000--000}
   \setcounter{page}{1}

   \author{Dipak Debnath\inst{1*}, Debjit Chatterjee\inst{1}, Arghajit Jana\inst{1,2}, Sandip K. Chakrabarti\inst{1}, Kaushik Chatterjee\inst{1}
}

   \institute{Indian Centre for Space Physics, 43 Chalantika, Garia St. Rd., Kolkata, 700084, India; {\it dipakcsp@gmail.com}\\
\and
   Physical Research Laboratory, Navrangpura, Ahmedabad 380009, India\\
%\vs \no
   {\small Received 2020 January 3; accepted 2020 April 5}
}

%\begin{abstract}
\abstract{
We study spectral and temporal properties of Galactic short orbital period transient black hole XTE~J1118+480 
during its 2005 outburst using archival data of RXTE PCA and HEXTE instruments in the combined energy range of $3-100$~keV.
Spectral analysis with the physical two-component advective flow (TCAF) model allows us to understand the accretion flow 
properties of the source. We found that this outburst of XTE~J1118+480 is an unconventional outburst as the source was 
only in the hard state (HS). Our spectral analysis suggests that during the entire outburst, the source was highly 
dominated by the low angular momentum sub-Keplerian halo rate. Since the source was active in radio throughout the outburst, 
we make an effort to estimate X-ray contribution of jets in total observed X-ray emissions from the spectral analysis with 
the TCAF model. The total X-ray intensity shows a similar nature of evolution as of radio and jet X-ray fluxes. This allowed 
us to define this `outburst' also as a jet dominated `outburst'. Total X-ray flux also found to subside when jet activity disappears. 
Our detailed spectral analysis also showed that although the source was only in the HS during the outburst, in the late declining 
phase spectrum became slightly softer due to the slow rise in the Keplerian disk rate.
%\end{abstract}
%\begin{keywords}
%\keywords{X-Rays:binaries -- stars individual:(XTE J1118+480) -- stars:black holes -- accretion, accretion disks -- ISM:
%\end{keywords}
}

\authorrunning{D. Debnath et al.}            %author_head in even pages
\titlerunning{Analysis of XTE J1118+480 Using TCAF Solution}  % title_head in odd pages
\maketitle

\section{Introduction}
XTE J1118+480 was discovered on 2000 March 29 by {\it All Sky Monitor} (ASM) onboard RXTE satellite at R.A.= $11^h18^m10^s.79$, 
Dec.= $48^\circ02'12''.42$ (Remillard et al. 2000). Follow up observations were carried out by Uemura et al., 2000; 
Chaty et al., 2000; Wren \& McKay, 2000; Mauche et al., 2000; Pooley \& Waldram, 2000. Because of its unique position 
in the Galactic halo, this black hole binary (BHB) suffered a very less absorption and is widely studied in multi-wavelength 
during both of its active outbursting periods and in the quiescence state (Revnivstev et al. 2000; Cook et al., 2000; Garcia et al., 2000;
Haswell et al., 2000; Hynes et al., 2000; McClintock et al., 2000; Wagner et al., 2000; Taranova et al., 2000; Esin et
al., 2001; Markoff et al., 2001; Hynes et al., 2003; Chaty et al., 2003; Shahbaz et al., 2005). The distance of this system 
is estimated to be around 1.8 kpc (McClintock et al. 2001a). The distance above the Galactic plane ($\sim1.7~kpc$) places 
it in the `Lockman halo' (Uemura et al. 2000). 
Quasi Periodic Oscillations (QPOs) were observed in X-ray, optical and EUV (Wood et al. 2000; Haswell et al. 2000).
The mass of the black hole (BH) has been predicted dynamically by Wagner et al. (2001) ($6.0-7.7~M_\odot$), Gelino et al. (2006)
($8.53\pm0.6~M_\odot$), Khargharia et al. (2013) ($6.9-8.2~M_\odot$).
This is suggested to be a high inclined ($\sim68-79^\circ$) (Khargharia et al. 2013), short orbital period ($\sim4.1~hrs$)
(Patterson et al. 2000; Gonzalez Hernandez et al. 2012) low mass binary system.
The spectral and the temporal properties of the source during its 2000 outburst have been studied by Chatterjee et al. (2019; hereafter Paper-I), 
using RXTE PCA and HEXTE combined data with the physical two component advective flow (TCAF) model (Chakrabarti \& Titarchuk 1995) to understand 
the accretion flow dynamics of the source. Estimation of X-ray contribution of the jet made from spectral analysis with the TCAF model 
suggests that the outburst was dominated by the jets/outflows. Monotonic evolution of low frequency QPOs is also studied. 
In Paper-I, the probable mass of the black hole (BH) is also estimated from the spectral analysis in the range 
of $6.25-7.49~M_\odot$ or $6.99^{+0.50}_{-0.74}~M_\odot$. 
%A similar mass range is also obtained from present study.
 
Roughly after five years of the quiescence phase on 2005 Jan., XTE~J1118+480 exhibited a new outbursting activity of shorter duration
($\sim 1.5$~months) and lower intensity flux compared to its earlier (2000) outburst. Multi-wavelength studies of the source
during this outburst is also carried out by many authors. Zurita et al. (2005a) reported an optical outburst on January 9, 2005 .
The outburst was also detected at X-ray and radio wavelengths (Remillard et al. 2005; Pooley 2005; Rupen et al. 2005).
The outburst faded rapidly, and reached near quiescence by the late February (Zurita et al. 2005b). The evolution of long-term
lightcurve and outburst properties had been discussed by Zurita et al. (2006). %Similar to the 2000 outburst this was
%also considered to be a jet dominated outburst (Hynes et al., 2006; Maitra et al., 2009; Brocksopp et al., 2010; Paper-I).

In the present {\it paper}, our goal is to study accretion flow properties of the source during its 2005 outburst. 
We wanted to check whether nature of the source is similar to other shorter orbital period harder or type-II transient 
low-mass X-ray binaries recently studied by our group (see, Debnath et al. 2017 and references therein). 
Here, we also make an effort to study properties of jets, particularly in X-ray band. A detailed spectral analysis is made 
using physical TCAF model (Chakrabarti \& Titarchuk 1995), which is based on transonic flows (Chakrabarti, 1996). 
%The spectral properties are studied in Chakrabarti (1997) for variation of flow parameters. 
This TCAF configuration consists of two types of flows, namely, high viscous, high angular momentum Keplerian 
flow along the equatorial plane and low viscous, low angular momentum sub-Keplerian flow enveloping the Keplerian disk. Close to 
the BH, axisymmetric shock forms due to the centrifugal barrier and the supersonic sub-Keplerian flow suddenly jumps to the subsonic 
branch. The post-shock region puffs-up and becomes hot due to slowing down of the matter at the shock location. This puffed-up region 
between the shock and the inner sonic point just outside of the horizon acts as the Compton cloud. This is known as the CENtrifugal barrier 
supported BOundary Layer or CENBOL. Thermal soft photons from the Keplerian disk becomes hard via repeated inverse-Compton scattering 
in the region. Although this generalized accretion flow model was introduced more than two decades ago, its recent implementation 
(after generation of model {\it fits} file using $\sim 10^6$ number of theoretical spectra produced by varying five model input parameters) 
%: %mass of the BH $M_{BH}$, Keplerian disk rate $\dot{m_d}$, sub-Keplerian halo rate $\dot{m_d}$, shock location $X_s$ and compression ratio $R$) 
as an additive table into HEASARC's spectral analysis software package XSPEC allowed one to get a more clear picture about the
flow dynamics of several BH sources (Debnath et al. 2014, 2015a,b, 2017; Mondal et al. 2014, 2016; Jana et al. 2016, 2020a;
Chatterjee et al. 2016, 2019; Molla et al. 2017). Masses of a few black hole candidates (BHCs) have been measured in a better
accuracy from normalization independent spectral analysis with the TCAF model (Molla et al. 2016, 2017; Chatterjee et al. 2016, 2019;
Jana et al. 2016, 2020a; Debnath et al. 2017; Bhattacharjee et al. 2017; Shang et al. 2019).
Theoretically, the coupling of disk and jet connection have been studied based on the transonic flow model by several authors (Chakrabarti
1999a,b; Chattopadhyay \& Das 2007; Aktar et al. 2015). The outflow rate from the inflowing accretion rate have been calculated using 
hydrodynamics of infalling and outgoing transonic flows (Chakrabarti 1999a,b; Das \& Chakrabarti 1999). Chakrabarti \& Mandal (2006)
studied the two component accretion flows in the presence of synchrotron. Estimation of the contribution of the jet X-ray fluxes (if present) 
and their properties are also studied with the TCAF model (Jana et al. 2017, 2020b; Paper-I). Even frequencies of the dominating 
type-C QPOs are also predicted from the TCAF model fitted shock parameters (Debnath et al. 2014; Chatterjee et al. 2016).

This {\it paper} is organized in the following way: in \S 2, we discuss the observation and data analysis procedures.
In \S 3, we present the results from our spectral analysis using two types of models. Evolution of X-ray contribution of jets/outflows
during the outburst is studied with that of radio fluxes. %We also estimate the probable range of mass for this compact object from the spectral results. 
Finally, in \S 4, we summarize the result and present briefly a comparison between this and the 2000 outburst.

\section{Observations and Data Analysis}
In 2005, although the outburst was reported on Jan. 9 by Zurita et al. (2005a), RXTE started monitoring it
five days later on a daily basis. To find the broad spectral information in $3-100$~keV band, we studied 21 observations
of combined data of RXTE PCA and HEXTE instruments during 2005 January 14 (MJD=53384.99) to January 25 (MJD=53395.59).
HEASARC's software package HeaSoft version HEADAS 6.16 and XSPEC version 12.8 had been used for our analysis.
We followed the methods as mentioned in Debnath et al. (2013, 2015a) for data reduction and analysis.

RXTE PCA lightcurve in the energy range of $2-15$~keV and $2-25$~keV are generated using the event mode data of maximum time
resolution of $125~\mu s$. The power density spectra (PDS) are generated using XRONOS task `powspec' on 0.01 sec binned lightcurves.
We used 1 sec time binned background subtracted lightcurves of the proportional counter unit 2 (PCU2) to calculate
average PCA count rate in 2-25 keV for each observation.

In general, PCU2 data of `standard 2' mode (FS4a* in the energy range $3-25$~keV) and HEXTE science mode data (FS52* in the
energy range of $20-100$~keV) of Cluster 0 or A are used for spectral analysis. For some observations due to low s/n, we
consider HEXTE data only in $20-40$~keV band. The background subtracted spectra was first fitted with a single power-law (PL) model,
verifying that no significant thermal disk blackbody component required. After that, we refitted all the spectra using the TCAF solution
based additive table {\it fits} file. A fixed hydrogen column density $N_H=1.3\times10^{20}~atoms~cm^{-2}$ is used for photoelectric
absorption model {\it phabs} (Garcia et al. 2000; McClintock et al. 2001b). A 0.5\% systematic error is considered during spectral
analysis with the PL as well as the TCAF model.

To fit a spectrum with the TCAF model, one requires to supply four flow parameters. These parameters are: two type of accretion
rates: (i) Keplerian disk rate ($\dot m_d$ in Eddington rate $\dot M_{Edd}$), (ii) sub-Keplerian halo rate ($\dot m_h$ in
$\dot M_{Edd}$), and two shock parameters : (iii) shock location ($X_s$ in Schwarzschild radius $r_s=2GM_{BH}/c^2$),
(iv) compression ratio ($R=\rho_+/\rho_-$, where $\rho_+$ and $\rho_-$ are the densities of post- and pre- shock regions respectively).
In addition, if the mass of the black hole ($M_{BH}$ in $M_\odot$) is known, one needs to provide the value of the mass and keep as frozen.
Otherwise, each fitted spectrum returns a value of the mass. One also gets a normalization ($N$) from each fitting. This normalization
parameter is a function of the mass of the black hole ($M_{BH}$), the distance ($D$) of the system, and the inclination angle ($i$) of the disk.
So for a specific black hole, this normalization value is expected to be a constant parameter, if measured with a given instrument
(Molla et al. 2016, 2017), excluding the facts that the system is not precessing and there is no significant jet/outflow from the disk.
Jana, Chakrabarti \& Debnath (2017; hereafter JCD17) developed a method to estimate X-ray contribution of jets/outflows from
spectral analysis with the TCAF model. They compared the variation of the model normalization with the observed radio fluxes, since
radio emission is known to be a tracer of jets/outflows. If there is a significant X-ray contribution from jets, we may require
a higher value of the normalization to fit a spectrum to take care of extra photon contributions from the base of the jets.
While comparing $N$ with the radio flux ($F_R$), if lower $N$ is found when $F_R$ is also at its minimum, then it implies
that during this low normalization days there is insignificant or no X-ray contribution from the jets in the net emission spectrum.
In other words, a large amount of outflow results in higher values of $N$. They also prescribed the procedure to calculate
the amount of the X-ray outflow flux ($F_{ouf}$). Following their method in Paper-I, Chatterjee et al. (2019) calculated jet X-ray fluxes
during the 2000 outburst of XTE J1118+480, where the minimum  value of $N$ is observed at $4.36$. Since the current outburst of XTE~J1118+480 
is also highly jet dominated, using the same method, we estimated the contribution of the jet X-ray flux in total flux for all observations.

Note: Since the main goal of this paper is not to measure the mass of the BH, we kept mass of the BH frozen at $7~M_\odot$ while fitting 
energy spectra with the TCAF model. This is the mean value of the TCAF model fitted mass values obtained from the spectral analysis 
of the 2000 outburst of XTE~J1118+480 (see, Paper-I).

\section{Results}
We study accretion properties of XTE J1118+480 during its 2005 outburst by analyzing $21$ RXTE PCA and HEXTE observations, selected 
from 2005 January 14 (MJD=53384.99) to January 25 (MJD=53395.59). Both temporal and spectral properties of the source are 
studied. Spectral study is done with the phenomenological (PL) and physical (TCAF) accretion flow models.

In Fig. 1, one example of power density spectrum (PDS) is shown along with the TCAF fitted combined PCA plus HEXTE spectrum of the 
data with observation ID 90011-01-01-04. In Fig. 2, variation of PCU2 count rate (in $2-25$~keV) and spectral parameters fitted 
with two different types of models: PL (PL flux, PL photon index $\Gamma$) and TCAF (Keplerian disk rate $\dot m_d$, sub-Keplerian 
halo rate $\dot m_h$, shock location $X_s$ and compression ratio $R$) are shown. 
Since we have been able to separate total observed X-ray into its two constituents: contribution from $i)$ inflowing matter or 
accretion disk, and $ii)$ outflowing matter or jets; in Fig. 3(a)-(c) we show variation of total X-ray flux ($F_{X}$), X-ray fluxes 
from accretion disk ($F_{inf}$) and jets ($F_{ouf}$). The variation of model normalization ($N$) and observed radio flux ($F_{R}$) 
are also plotted in Fig. 3(d)-(e). The radio flux ($F_R$ in mJy) data of $15$~GHz Ryle Telescope are shown here, are 
adopted from Brocksopp et al. (2010).
%In Fig. 5(a)-(b) we have plotted the contours of $\dot m_d$ vs. $\dot m_h$ and $X_s$ vs. $R$ for the first observation (MJD=53384.99).

\begin{figure}[h]
\vskip 0.1cm
        \centerline{
        \includegraphics[scale=0.6,width=7truecm,angle=0]{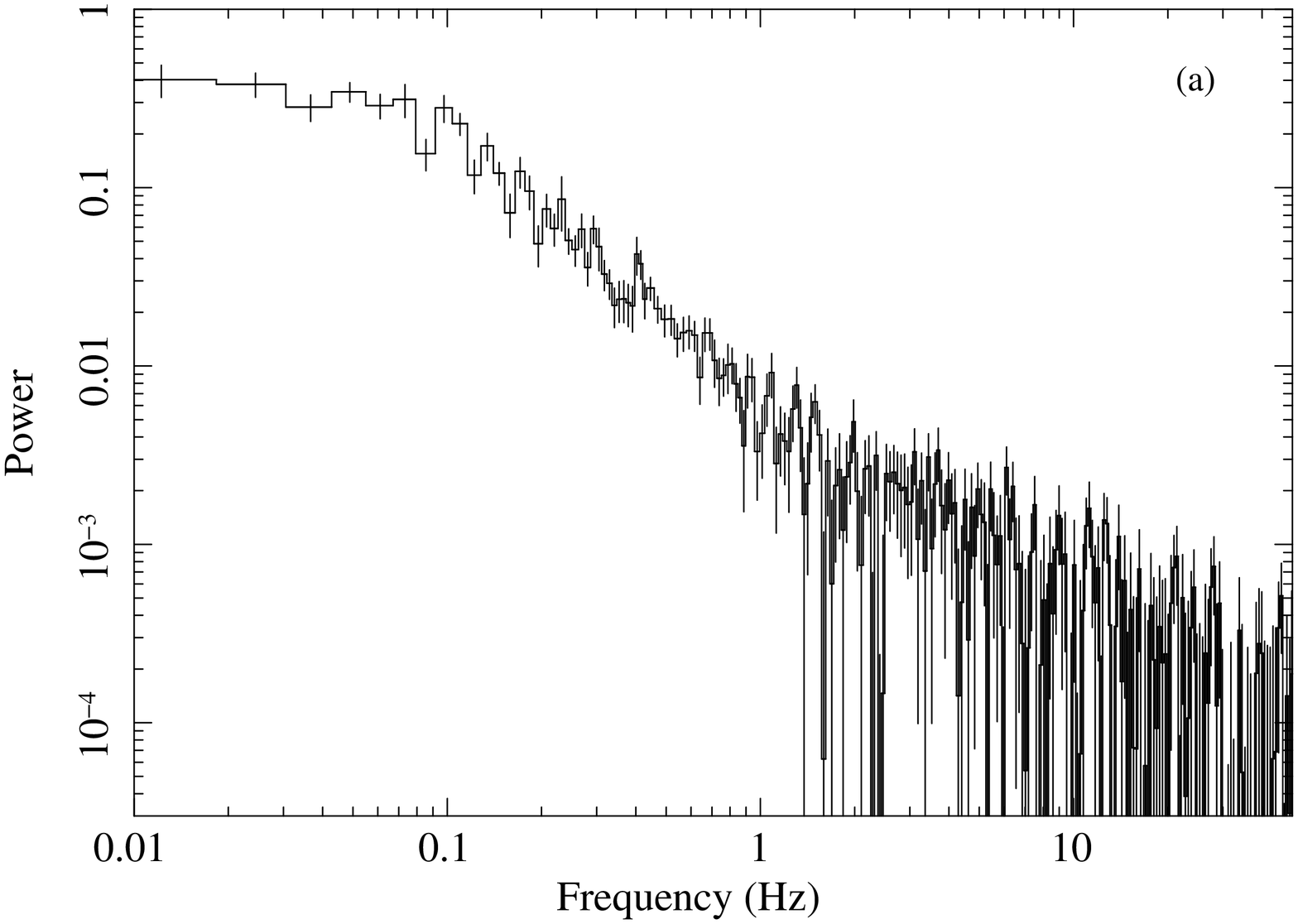}\hskip 0.3cm %}\vskip 0.1cm
%        \centerline{
        \includegraphics[scale=0.6,width=7truecm,angle=0]{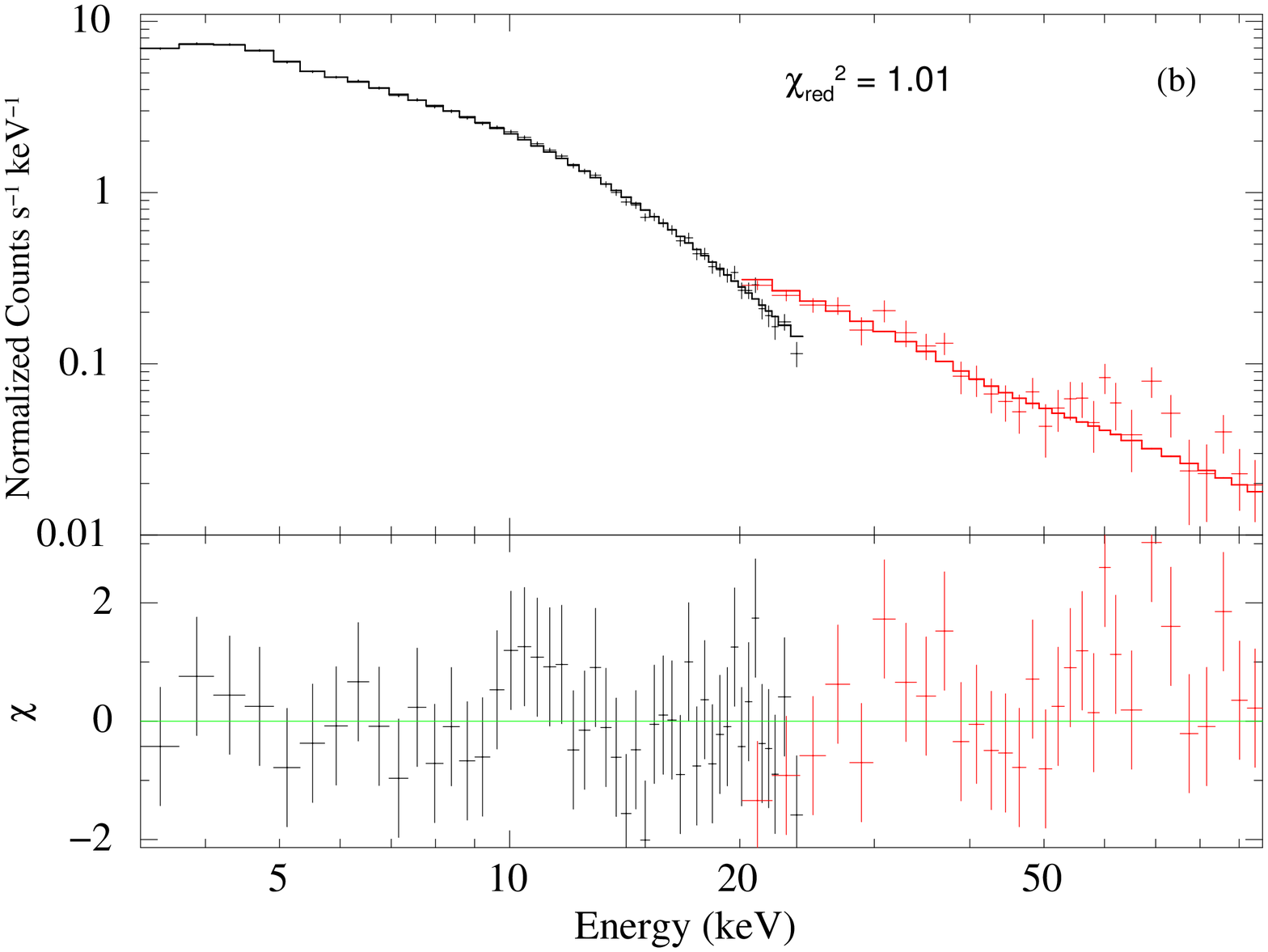}
        }
\caption{(a) A sample power density spectrum of $0.01$~sec time binned PCU2 lightcurve, and (b) TCAF model fitted 
combined PCA plus HEXTE spectrum in $3-100$~keV energy range for observation ID 90011-01-01-04 (MJD=53387.80) are shown.}
       \label{fig1}
\end{figure}

The detailed spectral analysis results are presented in Tables 1 \& 2. In Table 1, both PL and TCAF model fitted spectral 
parameters are presented. Different types of X-ray fluxes ($F_{X}$, $F_{inf}$, $F_{ouf}$), the percentage of the contribution 
of the jet X-ray in total X-ray, and TCAF model fitted normalization parameter values are presented in Table 2.

\subsection{Temporal Study}
RXTE PCA started to monitor the source five days later of the report of the outburst on 2005 Jan. 9 by Zurita et al. (2005a). 
Since this is a Fast Rise Slow Decay (FRSD) type outburst (see, Debnath et al. 2010), during the first PCA observation, the source 
already crossed its peak flux (see, Fig. 2(a)). It showed a monotonic  decrease in the PCU2 photon count rate in $2-25$~keV energy 
band. Fourier transformed PDSs (see, Fig. 1(a)) are generated for all observations using $0.01$~sec time binned $2-15$~keV lightcurves. 
We have not seen any prominent signature of the low frequency QPOs, which are generally observable during the outburst of 
this type of transient BHCs. Contrary to this, a monotonic evolution (increasing frequency from $0.06$ to $0.16$~Hz) of the 
QPOs were observed during 2000 outburst of the source (see, Paper-I). However, there is a report of the presence of a mHz QPO 
by Shahbaz et al. (2005) in the optical waveband on June 2003, which is the quiescent phase between 2000 and 2005 outbursts.

\begin{figure}[h]
\vskip 0.1cm
\centerline{
            \includegraphics[scale=0.6,width=14.0truecm,angle=0]{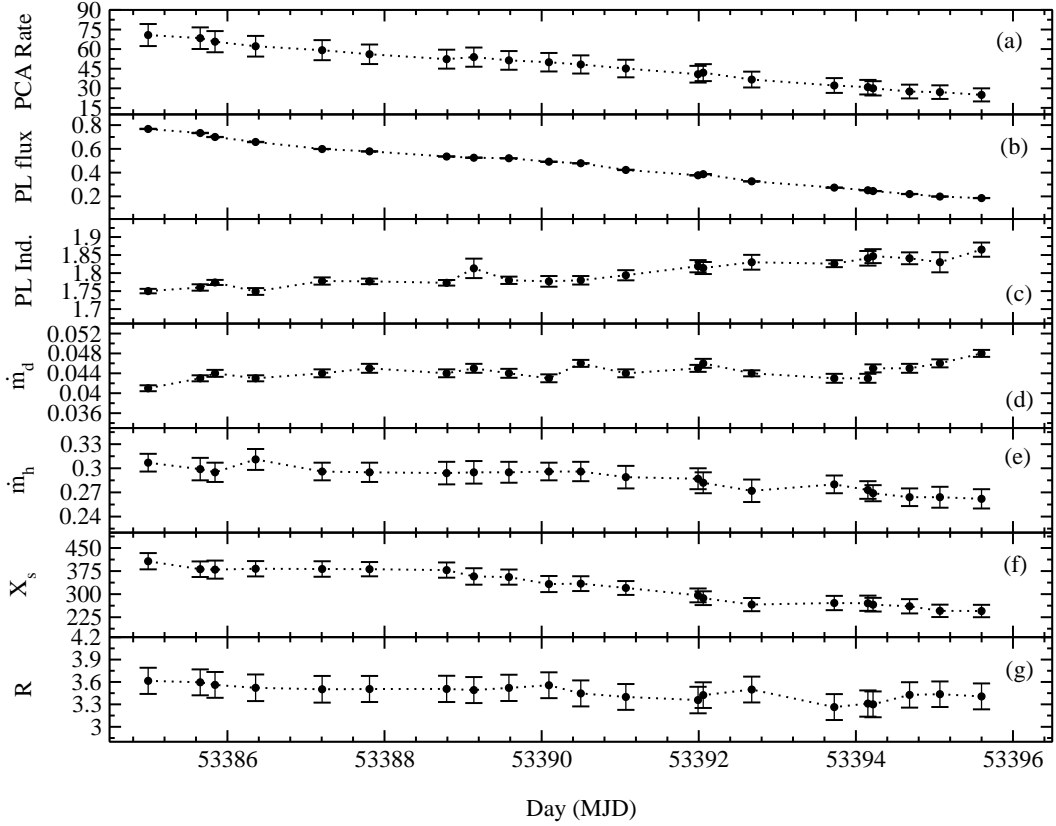}
           }
\caption{Variation of (a) PCU2 count rate (counts/sec) in $2-25$~keV energy range, PL model fitted (b) PL flux in unit of 
$10^{-9}~erg~cm^{-2}s^{-1}$, (c) PL photon index ($\Gamma$), TCAF model fitted (d) disk rate ($\dot{m}_d$) in $\dot{M}_{Edd}$ 
(e) halo rate ($\dot{m}_h$) in $\dot{M}_{Edd}$, (f) shock location ($X_s$) in $r_s$, (g) compression ratio ($R$) with day 
(in MJD) are plotted.}
\label{fig2}
\end{figure}

\subsection{Spectral Study}
The spectral analysis is done independently with the two types of models: $i)$ phenomenological PL model, and $ii)$ physical TCAF 
model. While PL model fit gives us an overview about the spectral properties of the source during different phases of the outburst, 
TCAF model fitted physical flow parameters allow us to get a the accretion flow dynamics and evolution of the flow geometry.

First, we fitted the background subtracted spectra solely with the PL model. No significant disk blackbody contribution 
was noticed during the spectral fittings. Although low values of the PL photon indices ($\Gamma \sim 1.75-1.87$) are 
observed during the outburst, in the late declining phase, a slow increase in $\Gamma$ values (from $1.81$ to $1.87$) are 
observed (see, Fig. 2(c)). On the first observation (MJD= 53384.99) day, $\Gamma=1.75$ is observed, and it varied in a 
narrow range of $\sim 1.75-1.81$ for the next seven days till MJD=53392.05 before rising slowly in the next three days 
to reach its maximum value of $1.87$ on the last day (MJD=53395.59) of our observation. The observation of the low $\Gamma$ 
signifies a hard spectral state, which is similar to what was seen during its earlier (2000) outburst of this source 
(Chaty et al. 2003; Paper-I). PL flux showed a similar nature of monotonic decrease in PCA count rate during the entire 
phase of the outburst (see Fig. 2(a) and 2(b)). 

To obtain a physical picture of the accretion flow dynamics of the source during its 2005 outburst, we refitted all spectra 
with the TCAF model based {\it fits} file as an additive table model in XSPEC. During spectral fit, we kept mass parameter 
frozen at $7~M_\odot$ as this was the estimated mass value obtained from spectral analysis with the TCAF model during 2000 
outburst of the source (Paper-I). Throughout the observation period, the Keplerian disk rate ($\dot m_d$) was observed to be 
very less as compared to the sub-Keplerian halo rate ($\dot m_h$). This result is consistent with the  non-requirement of the 
disk blackbody (DBB) component while fitting spectra with the PL model. On the first observation (MJD=53384.99) day, the value 
of $\dot m_d$ was $0.041$ $\dot M_{Edd}$. On the second observation day, it increased slightly ($\sim 0.044$ $\dot M_{Edd}$), 
and subsequently became roughly constant in that value except during the last few observations when a slow monotonic increase 
(up to $0.48$~$\dot M_{Edd}$) was observed. The sub-Keplerian halo rate ($\dot m_h$) showed a monotonic decrease from $0.31$ to 
$0.26$ $\dot M_{Edd}$ during the entire period of the outburst. The variations of $\dot m_d$ and $\dot m_h$ are plotted in 
Fig. 2(d)-(e). The variation of the TCAF model fitted shock parameters (location $X_s$ and compression ratio $R$) are shown 
in Fig. 2(f)-(g). On the first observation day, a strong shock ($R=3.61$) at a far location ($X_s=406~r_s$) was observed. 
As the day progresses, shock becomes weaker and moves inward. On the last observation day, a comparatively weaker shock 
($R=3.40$) at $X_s=245~r_s$ is observed.
%A sample contour plot of $\dot m_d$ vs. $\dot m_h$ and $X_s$ vs. $R$ for the first observation (MJD=53384.99) are shown in Fig. 5(a)-(b). 

\subsection{Jet X-ray}
Radio emission observed in BHCs is considered to be a tracer of jets and outflows. The strong radio emission is reported during 
most part of the outburst (see, Hynes et al., 2006; Maitra et al., 2009; Brocksopp et al. 2010). During the 2000 outburst, 
similar strong radio emission is observed (see, http://www.mrao.cam.ac.uk/$\sim$guy/J1118+480/J1118480.list). It is to be noted
that the radio flux ($F_R$) decreases as PCA rate decreases (see, Fig. 3(e) and Fig. 2(a)). So, it appears that the outburst 
is totally dominated by the emission from jets/outflows. This motivated us to separate X-ray contribution from the jet/outflow 
($F_{ouf}$) from that of the accretion disk or inflowing matter ($F_{inf}$) using the same method as described in JCD17 and Paper-I. 

According to TCAF, model normalization parameter is a function of the mass of the black hole ($M_{BH}$), the inclination 
angle ($i$) and the distance ($D$) of the system. So when there is no prominent X-ray contribution from other physical 
processes (whose effects are not considered in the current version of the TCAF model fits file), for example, jets or outflows 
or precession in the disk, $N$ should not vary from one day to another if observed with a given instrument. But during the outburst, 
we see a monotonic decrease in $N$ from $10.9$ to $4.41$ (see, Fig. 3(d)). Its variation is roughly similar to $F_R$ (see Fig. 3(e)). 
This allows us to conclude that higher $N$ values are possibly required to fit the spectra to match excess contribution of X-rays, 
emitted from the base of the jet.

Total X-ray fluxes ($F_X$) are obtained in our PCA spectral analysis band ($3-25$~keV), when all the TCAF model input 
parameters (except the mass of the BH) are kept free. In presence of jet or outflow, $F_X$ is the combined contribution of 
the radiations from the accretion disk and CENBOL, i.e. from inflowing matter ($F_{inf}$) and from the base of the jet i.e., from 
outflowing matter ($F_{ouf}$). Significant X-ray contribution from the jet to the total X-ray influences our spectral fitting 
with the TCAF model. Since present version of the TCAF model fits file takes care only the radiation contributed from the 
inflowing matter, higher values of normalization are required to fit the spectra as to compensate the excess X-ray radiations 
emitted from the jet. In the absence of jet, an almost constant value of model normalization is sufficient to fit the spectra 
during the entire outburst (see Molla et al. 2016, 2017; Chatterjee et al. 2016). For the case of 2005 outburst of XTE J1118+480, 
we looked into the obtained $N$ values and find that on the last data (MJD=53395.59), the model normalization was at its minimum 
value of $4.41$. This implies that on this observation total X-ray flux was contributed only by the radiations from the accretion 
disk and CENBOL. So, we may say that on this observation jets X-ray was minimum or negligible. This was indeed observed 
via variation of the radio fluxes (see, Fig. 3(e)). So, after confirmation from low $F_R$ at the lowest $N$ observed day, 
we may obtain $F_{inf}$ during the entire outburst by refitting spectra with the frozen model $N$ value (at its lowest 
observed value, obtained when all model flow parameters are kept free). Now we could separate jet component of the X-ray 
fluxes ($F_{ouf}$) in each observations by subtracting $F_{inf}$ from $F_X$ (observed flux when all model flow parameters 
are kept free), i.e., 
%\begin{equation}
$$
F_{ouf}=F_X - F_{inf}  \eqno(1)
$$
%\end{equation}
This method of separating jet contribution of X-ray from total observed X-ray was introduced in Jana et al. (2017).

Although during 2005 outburst of XTE~J1118+480, minimum $N$ value $4.41$ was required to fit spectra, according to Paper-I, 
lowest $N$ value was required to fit 2000 outburst spectra was $4.37$. Since Paper-I $N$ value was lowest between two 
outbursts, to obtain X-ray contribution only from the disk and CENBOL i.e., inflowing matter, we refitted all 2005 studied 
PCA spectra with the TCAF model after keeping model normalization frozen at $4.37$. Finally, $F_{ouf}$ is calculated for each 
observation by taking the difference of $F_{inf}$ and $F_X$ as in Eqn (1). 
The evolution of $F_X$, $F_{inf}$, $F_{ouf}$, $N$, and $F_R$ are shown in Fig. 3. $F_{ouf}$ shows a similar 
variation as of $N$ and $F_R$. When we see the percentage of jet X-ray contribution in total emitted X-ray, as 
outburst progresses it decreases (see, Col. 6 of Table 2). Maximum fractional jet X-ray contribution is found 
to be $\sim 60\%$.

\begin{figure}[h]
\vskip 0.2cm
\centerline{
           \includegraphics[scale=0.6,width=14.0truecm,angle=0]{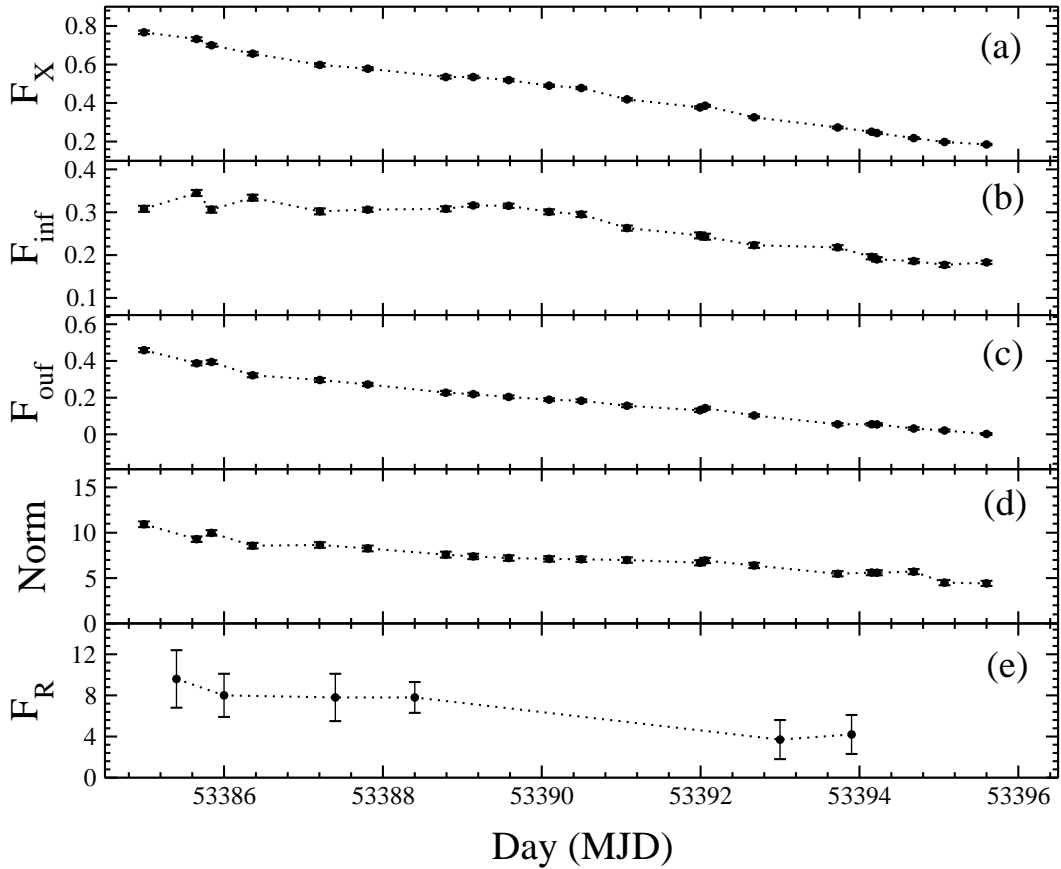}
           }
\caption{Variation of (a) total flux ($F_X$), (b) inflow or disk flux ($F_{inf}$), (c) outflow or jet flux ($F_{ouf}$), 
(d) model normalization ($N$), and (e) radio flux ($F_R$) are shown with MJD. $F_X$, $F_{inf}$, $F_{ouf}$ are obtained 
TCAF model fitted spectra in $3-25$~keV PCA range in units of $10^{-9}~erg~cm^{-2}~s^{-1}$ unit. $F_R$ of $15~GHz$ Ryle 
Telescope is presented in $mJy$ unit and adopted from Brocksopp et al. (2010).} 
\label{fig4}
\end{figure}

\section{Summary and Discussions}

The 2005 outburst of XTE J1118+480 (epoch under consideration here) is a shorter duration, lower intense and less studied 
compared to its earlier 2000 outburst. To understand the nature of the accretion flow dynamics of the source during this outburst, 
we use RXTE PCA and HEXTE combined data for our temporal and spectral analysis. We did not find any prominent signature of 
low frequency QPOs in the $0.01$~sec binned Fourier transformed power density spectra, although Shahbaz et al. (2005) reported 
a QPO at $\sim 2$~mHz from the optical data in the quiescent state (June 2003), which is in between two outbursts of the source. 
The spectral analysis is done using two types of models: the phenomenological power-law (PL) model and physical TCAF solution 
based {\it fits} file. Total $21$ observed data of combined PCA and HEXTE spectra in the energy range of $3-100$~keV 
(except some observations where due to low s/n, $3-40$~keV data are used) are chosen for spectral fittings from 2005 January 14 
(MJD=53384.99) to January 25 (MJD=53395.59). Although triggering of the outburst was reported by Zurita et al. (2005a) on 
2005 Jan. 9, RXTE started monitoring it five days later. So we missed an important rapidly evolving rising phase of the outburst. 

The evolution of the average PCA rate in $2-25$~keV band implies that on the first observation day (MJD=53384.99), the source 
already passed its peak intensity (see, Fig. 2(a)). During the entire period of the outburst, the photon count was found to be 
monotonically decreasing. While fitting spectra with the PL model, we noticed the same behaviour for the PL flux (see Fig. 2(b)). 
We tried to fit the spectra with combined DBB plus PL models, but `ftest' suggest that DBB component is insignificant. This 
means that during the entire outburst, non-thermal PL photons are highly dominating over the low or insignificant thermal disk 
component. PL model fitted photon index ($\Gamma$) values are obtained in a lower range ($\sim1.75-1.87$), implies that during 
the entire phase of the outburst, the source was in the hard state. If we relook at the $\Gamma$ values, we see an increasing 
trend of it (from $1.81$ to $1.87$) i.e., rise in softness in the late declining phase of the outburst. There is no physical 
explanation of this $\Gamma$ from the PL model fitted spectral analysis.

During the entire outburst, high dominance of the sub-Keplerian halo rate ($\dot m_h$) over the Keplerian disk rate ($\dot m_d$) 
in the presence of strong shock ($R>3$) far away from the BH ($X_s>245~r_s$) are observed. But if we look into the evolution 
of accretion rates, in the late declining phase (specifically in last observations) a slow rise in $\dot m_d$ is observed although 
$\dot m_h$ is decreased. This explains the gradual softening i.e., increase in $\Gamma$ in the late declining phase of the outburst. 
This feature is quite uncommon in an outburst of a transient BHC. This was due to the vanishing of the jet effect.
A similar late softening was also observed due to the slow rise in $\dot m_d$ during the late declining phase of 2000 outburst 
(Paper-I). So the source has been consistently behaving in a non-conventional way even after a gap of five long years. 
This points to the fundamental configuration of the binary and compactness of the system.

Radio flares are believed to be a tracer of jets or outflows. Similar to the 2000 outburst, high radio fluxes are also observed 
during the current outburst of XTE~J1118+480. $F_R$ showed similar declining variation as of PCA rate, PL flux, $F_X$. 
Now, when we looked at the TCAF model fitted $N$ values, similar nature of monotonically decreasing values (from $10.9$ 
to $4.41$) are observed. This means that higher $N$ were required to fit spectra when stronger jets were present.
We estimated $F_{inf}$ by refitting the spectra with $N$ frozen at its minimum observed value ($=4.37$) during its 2000 and 2005 
outbursts (when spectral fits are done with all TCAF model parameters as free).
We estimate the X-ray flux contribution from jets ($F_{ouf}$) and find that the fractional jet X-ray contribution ($F_{ouf}$/$F_X$) 
is maximum $\sim 60\%$. Evolution of $F_{ouf}$ looks similar to $F_R$. It is to note that the outburst is suppressed with the 
decrease in $F_{ouf}$. Therefore, similar to the 2000 outburst of XTE~J118+480, this so-called `failed outburst' could be termed as 
jet activity dominated outburst.

%####
\section*{Acknowledgements}
This work made use of PCA and HEXTE data of NASA's RXTE satellite.
Research of D.D. and S.K.C. is supported in part by the Higher Education Dept. of the Govt. of West Bengal, India. 
D.D. and S.K.C. also acknowledge partial support from ISRO sponsored RESPOND project (ISRO/RES/2/418/17-18) fund.
D.C. and D.D. acknowledge support from DST/SERB sponsored Extra Mural Research project (EMR/2016/003918) fund.
A.J. and D.D. acknowledge support from DST/GITA sponsored India-Taiwan collaborative project (GITA/DST/TWN/P-76/2017) fund.
A.J.  acknowledges CSIR SRF fellowship (09/904(0012)2K18 EMR-1).
K.C. acknowledges support from DST/INSPIRE fellowship (IF170233).

%\clearpage
\bibliographystyle{raa}

%%%%%%%%%%%%%%%%%%%%%%%%%%%%%%%%%%%%%%%%%%%%%%%%%%

%\clearpage

%\clearpage
%\clearpage

\clearpage
\begin{table}
\small
\addtolength{\tabcolsep}{-3.0pt}
%\centering
\caption{PL or TCAF model fitted spectral parameters}
\label{tab:table1}
\begin{tabular}{lcc|ccc|cccccc}
\hline
Obs. ID& UT    & MJD & PL Ind       & PL flux & $\chi^2/dof$ & $\dot{m}_d$       & $\dot{m}_h$       & $X_s$   & R  &  $\chi^2/dof$ \\
       & Date  &   &  ($\Gamma$)  &         &              & ($\dot{M}_{Edd}$) & ($\dot{M}_{Edd}$) & $(r_s)$ &    &               \\
 (1)   & (2)   &   (3)        &  (4)    &   (5)        &  (6)              &      (7)          &  (8)    &(9)&  (10)   &  (11)  \\
\hline

X-01-00& 14/01& 53384.99& $1.750^{\pm0.006}$ &$0.768^{\pm0.002}$  &45.5/47&$0.041^{\pm0.0006}$& $0.30^{\pm0.011}$& $406^{\pm26}$& $3.61^{\pm0.17}$&  89.4/68\\
X-01-07& 15/01& 53385.65& $1.760^{\pm0.008}$ &$0.734^{\pm0.002}$  &39.2/47&$0.043^{\pm0.0006}$& $0.29^{\pm0.014}$& $381^{\pm25}$& $3.59^{\pm0.17}$&  65.2/68\\
X-01-02& 15/01& 53385.84& $1.774^{\pm0.007}$ &$0.700^{\pm0.001}$  &46.6/47&$0.044^{\pm0.0007}$& $0.29^{\pm0.012}$& $379^{\pm29}$& $3.56^{\pm0.17}$&  68.0/71\\
X-01-08& 16/01& 53386.35& $1.749^{\pm0.009}$ &$0.658^{\pm0.002}$  &44.1/47&$0.043^{\pm0.0006}$& $0.31^{\pm0.013}$& $382^{\pm24}$& $3.52^{\pm0.17}$&  69.5/71\\
X-01-09& 17/01& 53387.20& $1.778^{\pm0.010}$ &$0.599^{\pm0.001}$  &41.3/47&$0.044^{\pm0.0008}$& $0.29^{\pm0.011}$& $381^{\pm25}$& $3.50^{\pm0.17}$&  62.4/71\\
X-01-04& 17/01& 53387.80& $1.777^{\pm0.007}$ &$0.579^{\pm0.001}$  &43.4/47&$0.045^{\pm0.0009}$& $0.29^{\pm0.012}$& $381^{\pm23}$& $3.50^{\pm0.17}$&  68.7/68\\
X-01-05& 18/01& 53388.79& $1.773^{\pm0.007}$ &$0.536^{\pm0.001}$  &69.7/47&$0.044^{\pm0.0008}$& $0.29^{\pm0.014}$& $378^{\pm24}$& $3.50^{\pm0.17}$&  97.4/71\\
X-01-11& 19/01& 53389.13& $1.813^{\pm0.027}$ &$0.526^{\pm0.002}$  &49.2/47&$0.045^{\pm0.0009}$& $0.29^{\pm0.014}$& $357^{\pm26}$& $3.49^{\pm0.17}$&  44.9/40\\
X-01-12& 19/01& 53389.58& $1.780^{\pm0.010}$ &$0.521^{\pm0.001}$  &36.7/47&$0.044^{\pm0.0009}$& $0.29^{\pm0.013}$& $355^{\pm24}$& $3.52^{\pm0.17}$&  42.1/51\\
X-01-13& 20/01& 53390.09& $1.777^{\pm0.014}$ &$0.492^{\pm0.002}$  &54.2/47&$0.043^{\pm0.0008}$& $0.29^{\pm0.011}$& $332^{\pm26}$& $3.55^{\pm0.17}$&  62.4/51\\
X-01-06& 20/01& 53390.49& $1.780^{\pm0.011}$ &$0.479^{\pm0.001}$  &39.0/47&$0.046^{\pm0.0007}$& $0.29^{\pm0.012}$& $334^{\pm24}$& $3.44^{\pm0.17}$&  46.7/46\\
Y-02-00& 21/01& 53391.07& $1.794^{\pm0.014}$ &$0.423^{\pm0.002}$  &43.9/47&$0.044^{\pm0.0008}$& $0.28^{\pm0.014}$& $319^{\pm22}$& $3.40^{\pm0.17}$&  43.7/48\\
Y-02-02& 21/01& 53391.99& $1.819^{\pm0.017}$ &$0.378^{\pm0.002}$  &34.9/47&$0.045^{\pm0.0007}$& $0.28^{\pm0.013}$& $295^{\pm22}$& $3.35^{\pm0.17}$&  36.6/46\\
Y-02-03& 22/01& 53392.05& $1.814^{\pm0.016}$ &$0.387^{\pm0.002}$  &25.4/47&$0.046^{\pm0.0009}$& $0.28^{\pm0.013}$& $286^{\pm22}$& $3.42^{\pm0.17}$&  27.9/46\\
Y-02-04& 22/01& 53392.67& $1.830^{\pm0.020}$ &$0.327^{\pm0.001}$  &48.9/47&$0.044^{\pm0.0006}$& $0.27^{\pm0.014}$& $265^{\pm21}$& $3.49^{\pm0.17}$&  56.2/46\\
Y-02-06& 23/01& 53393.72& $1.826^{\pm0.009}$ &$0.274^{\pm0.001}$  &51.3/47&$0.043^{\pm0.0009}$& $0.28^{\pm0.011}$& $271^{\pm23}$& $3.26^{\pm0.17}$&  121/71\\
Y-02-14& 24/01& 53394.15& $1.841^{\pm0.020}$ &$0.252^{\pm0.001}$  &29.6/47&$0.043^{\pm0.0009}$& $0.27^{\pm0.011}$& $270^{\pm24}$& $3.31^{\pm0.17}$&  34.2/46\\
Y-02-15& 24/01& 53394.21& $1.847^{\pm0.019}$ &$0.245^{\pm0.001}$  &38.5/47&$0.045^{\pm0.0008}$& $0.26^{\pm0.010}$& $265^{\pm22}$& $3.30^{\pm0.17}$&  44.0/46\\
Y-02-07& 24/01& 53394.68& $1.841^{\pm0.016}$ &$0.220^{\pm0.001}$  &44.0/47&$0.045^{\pm0.0009}$& $0.26^{\pm0.011}$& $260^{\pm23}$& $3.42^{\pm0.17}$&  33.6/46\\
Y-02-09& 25/01& 53395.06& $1.830^{\pm0.027}$ &$0.199^{\pm0.001}$  &42.5/47&$0.046^{\pm0.0008}$& $0.26^{\pm0.013}$& $245^{\pm20}$& $3.43^{\pm0.17}$&  47.9/46\\
Y-02-10& 25/01& 53395.59& $1.865^{\pm0.019}$ &$0.186^{\pm0.001}$  &26.9/47&$0.048^{\pm0.0007}$& $0.26^{\pm0.012}$& $245^{\pm20}$& $3.40^{\pm0.17}$&  26.1/43\\

\hline
\end{tabular}
\noindent{
\leftline{X=90011-01, Y=90111-01 are prefixes of observation IDs. UT date is in dd/mm format of year 2005.}
\leftline{$\Gamma$ represents the photon indices obtained from pure PL model fitting. PL flux indicates the flux}
\leftline{from PL model in $10^{-9}~erg~cm^{-2}s^{-1}$. $\dot{m}_d$, $\dot{m}_h$, $X_s$, $R$ are the}
\leftline{TCAF fitted parameters. The accretion rates ($\dot{m}_d$ and $\dot{m}_h$) are in Eddington rate.}
\leftline{$X_s$ is the shock location values in $r_s$ unit and $R$ is the compression ratio. PL and TCAF model fitted $\chi^2_{red}$ values} 
\leftline{are mentioned as $\chi^2$/dof in Cols. 6 \& 11 respectively, where `dof' represents the degrees of freedom.}
\leftline{The superscripts are average error values of $\pm$ 90\% confidence extracted using `err' task in XSPEC.}
}
\end{table}

\clearpage
\begin{table}
%\small
\addtolength{\tabcolsep}{-3.0pt}
%\centering
\caption{X-ray Flux Contributions of Total, Accretion disk, and Jets}
\label{tab:table2}
\begin{tabular}{lc|ccc|c|c}
\hline
Obs Id. & MJD & $F_X$ & $F_{inf}$ & $F_{ouf}$ & \% of $F_{ouf}$& Norm  \\
(1)     &(2)   & (3) &  (4)  & (5)       &  (6)      &  (7)      \\
\hline
X-01-00& 53384.99& $0.767^{\pm0.009}$ &$0.308^{\pm0.007}$ &$0.459^{\pm0.011}$ &$59.84^{\pm1.59}$& $10.93^{\pm0.312}$\\
X-01-07& 53385.65& $0.732^{\pm0.009}$ &$0.345^{\pm0.007}$ &$0.387^{\pm0.011}$ &$52.86^{\pm1.63}$& $9.287^{\pm0.299}$\\
X-01-02& 53385.84& $0.700^{\pm0.008}$ &$0.306^{\pm0.007}$ &$0.394^{\pm0.010}$ &$56.28^{\pm1.56}$& $9.981^{\pm0.301}$\\
X-01-08& 53386.35& $0.656^{\pm0.009}$ &$0.334^{\pm0.007}$ &$0.322^{\pm0.011}$ &$49.08^{\pm1.80}$& $8.572^{\pm0.309}$\\
X-01-09& 53387.20& $0.598^{\pm0.009}$ &$0.302^{\pm0.007}$ &$0.296^{\pm0.011}$ &$49.49^{\pm1.98}$& $8.650^{\pm0.311}$\\
X-01-04& 53387.80& $0.578^{\pm0.008}$ &$0.306^{\pm0.005}$ &$0.272^{\pm0.009}$ &$47.05^{\pm1.68}$& $8.267^{\pm0.312}$\\
X-01-05& 53388.79& $0.535^{\pm0.008}$ &$0.308^{\pm0.006}$ &$0.227^{\pm0.010}$ &$42.42^{\pm1.97}$& $7.581^{\pm0.325}$\\
X-01-11& 53389.13& $0.535^{\pm0.007}$ &$0.316^{\pm0.004}$ &$0.219^{\pm0.008}$ &$40.93^{\pm1.58}$& $7.387^{\pm0.297}$\\
X-01-12& 53389.58& $0.519^{\pm0.007}$ &$0.315^{\pm0.005}$ &$0.204^{\pm0.008}$ &$39.30^{\pm1.63}$& $7.209^{\pm0.300}$\\
X-01-13& 53390.09& $0.490^{\pm0.006}$ &$0.301^{\pm0.006}$ &$0.189^{\pm0.008}$ &$38.57^{\pm1.69}$& $7.115^{\pm0.297}$\\
X-01-06& 53390.49& $0.478^{\pm0.006}$ &$0.295^{\pm0.006}$ &$0.183^{\pm0.008}$ &$38.28^{\pm1.74}$& $7.075^{\pm0.294}$\\
Y-02-00& 53391.07& $0.419^{\pm0.006}$ &$0.263^{\pm0.006}$ &$0.156^{\pm0.008}$ &$37.23^{\pm1.98}$& $6.992^{\pm0.297}$\\
Y-02-02& 53391.99& $0.377^{\pm0.006}$ &$0.246^{\pm0.007}$ &$0.131^{\pm0.009}$ &$34.74^{\pm2.45}$& $6.686^{\pm0.289}$\\
Y-02-03& 53392.05& $0.386^{\pm0.005}$ &$0.243^{\pm0.007}$ &$0.143^{\pm0.008}$ &$37.04^{\pm2.12}$& $6.940^{\pm0.292}$\\
Y-02-04& 53392.67& $0.326^{\pm0.005}$ &$0.223^{\pm0.006}$ &$0.103^{\pm0.007}$ &$31.59^{\pm2.20}$& $6.392^{\pm0.294}$\\
Y-02-06& 53393.72& $0.273^{\pm0.005}$ &$0.218^{\pm0.005}$ &$0.055^{\pm0.007}$ &$20.14^{\pm2.59}$& $5.472^{\pm0.284}$\\
Y-02-14& 53394.15& $0.251^{\pm0.005}$ &$0.196^{\pm0.006}$ &$0.055^{\pm0.007}$ &$21.91^{\pm2.82}$& $5.608^{\pm0.286}$\\
Y-02-15& 53394.21& $0.244^{\pm0.005}$ &$0.190^{\pm0.005}$ &$0.054^{\pm0.007}$ &$22.13^{\pm2.90}$& $5.593^{\pm0.281}$\\
Y-02-07& 53394.68& $0.218^{\pm0.005}$ &$0.186^{\pm0.005}$ &$0.032^{\pm0.007}$ &$23.39^{\pm3.25}$& $5.133^{\pm0.280}$\\
Y-02-09& 53395.06& $0.198^{\pm0.004}$ &$0.177^{\pm0.005}$ &$0.021^{\pm0.006}$ &$10.60^{\pm3.03}$& $4.498^{\pm0.290}$\\
Y-02-10& 53395.59& $0.185^{\pm0.004}$ &$0.183^{\pm0.004}$ &$0.002^{\pm0.005}$ &$1.081^{\pm2.70}$& $4.413^{\pm0.283}$\\
\hline
\end{tabular}
\noindent{
\leftline{X=90011-01, Y=90111-01 are prefixes of observation IDs.} 
\leftline{Total ($F_X$), accretion disk ($F_{inf}$), and Jet ($F_{ouf}$) X-ray fluxes are in units of}
\leftline{$10^{-9}~ergs~cm^{-2}~s^{-1}$ and they are calculated in $3-25$~keV PCA band.}
\leftline{TCAF model fitted normalization ($N$) values with errors are shown in Col. 8.}
\leftline{Note: average values of 90\% confidence $\pm$ values obtained using `err' task in XSPEC,}
\leftline{are plotted as superscripts of fitted parameter values.}
}
\end{table}

%\label{lastpage}
\end{document}